\documentclass[reprint]{revtex4-1}
\usepackage{amsmath}
\usepackage[pdftex]{graphicx} 
\usepackage{dcolumn}
\usepackage{bm}

\newcommand{\rem}[1]{ }
\newcommand{\beq}{\begin{equation}}
\newcommand{\eeq}{\end{equation}}
\newcommand{\bea}{\begin{eqnarray}}
\newcommand{\eea}{\end{eqnarray}}

\begin{document}
\title{Asymmetric diffusion of cosmic rays}

\author{Mikhail V. Medvedev} 
\affiliation{Department of Physics and Astronomy, University of Kansas, Lawrence, KS 66045}

\author{Viktor V. Medvedev}
\affiliation{K.G. Razumovskiy Moscow State University of Technologies and Management (First Cossack University), \\
73 Zemlyanoy Val, Moscow, Russia, 109004 }

\begin{abstract}
Cosmic ray propagation is diffusive because of pitch angle scattering by waves. We demonstrate that if the high-amplitude magnetohydrodynamic turbulence with $\tilde B/\langle B\rangle \sim 1$ is present on top of the mean field gradient, the diffusion becomes asymmetric. As an example, we consider the vertical transport of cosmic rays in our Galaxy propagating away from a point-like source. We solve this diffusion problem analytically using a one-dimensional Markov chain analysis. We obtained that the cosmic ray density markedly differs from the standard diffusion prediction and has a sizable effect on their distribution throughout the galaxy. The equation for the continuous limit is also derived, which shows limitations of the convection-diffusion equation. 
\end{abstract}
\keywords{cosmic rays: propagation}

\maketitle
 
\section{Introduction}

Propagation of charged particles, whose Larmor radii, $r_L$, are much smaller than the field inhomogeneities is ballistic along the field lines, neglecting drifts. When small-scale turbulence is also present, it can induce pitch-angle diffusion, so the parallel propagation becomes diffusive, and the particles can also jump across fields lines through a distance $\sim r_L$. It is this type of propagation of low energy cosmic rays (CR) through the Galaxy that is believed to occur. Indeed, the characteristic field correlation scale, $\lambda$, of the galactic magnetic field is of the order of a few parsecs \cite{32,39}. Hence, $r_L\ll \lambda$ for the CR below the knee, $E\lesssim$~PeV, so they nearly follow field lines in a diffusive manner. The small-scale Alfvenic turbulence responsible for the pitch-angle diffusion is believed to be self-generated by the streaming cosmic rays \cite{cr-turb}. 

The presence of strong magnetohydrodynamic (MHD) turbulence in the interstellar medium (ISM) \cite{32,39} also results in chaotic distribution of the filed lines in the galaxy. Such a chaotic field line topology leads to an additional, very efficient three-dimensional (3D) diffusion of CR throughout the entire ISM \cite{034,35}. Moreover, the MHD (likely Alfvenic) turbulence in the ISM is known to be of high amplitude at the outer scale, that is $\tilde B/\langle B\rangle \gtrsim 1$. Such high-amplitude magnetic field fluctuations affect diffusion via mirroring and transient trapping effects as well \cite{35,36,ptuskin-traps}. In addition, trapping in MHD turbulence is intermittent and transient because large-amplitude, quasi-coherent Alfv\'enic wave-forms (``magnetic traps'' or ``magnetic bottles'') are not static but exist for a certain life-time --- the Alfvenic time. 

Finally, the mean field in the Milky Way (and all other galaxies too) is non-uniform. There is a net gradient of $B$ toward the galactic mid-plane and the center of the galaxy. The exponential scale-height, $H$, of the galactic disk region dominated by magnetic fields is about a kiloparsec \cite{39}. Interestingly, particle diffusion with magnetic trapping in the presence of the net field gradient has, to our knowledge, never been addressed before, except for our earlier paper, where we argued for such a possibility \cite{MM07}. In this paper we demonstrate that the mean field gradient modifies diffusion drastically, so that it becomes asymmetric in which transition probabilities from a current particle's position onto neighboring ones are unequal. We stress that asymmetric diffusion is not to be confused with the standard anisotropic diffusion, in which the probabilities are equal but the diffusion coefficient can be inhomogeneous and anisotropic, in general.

\section{Transport in a ``multiple-mirror machine''}

\begin{figure}[b]
\includegraphics[scale=0.3]{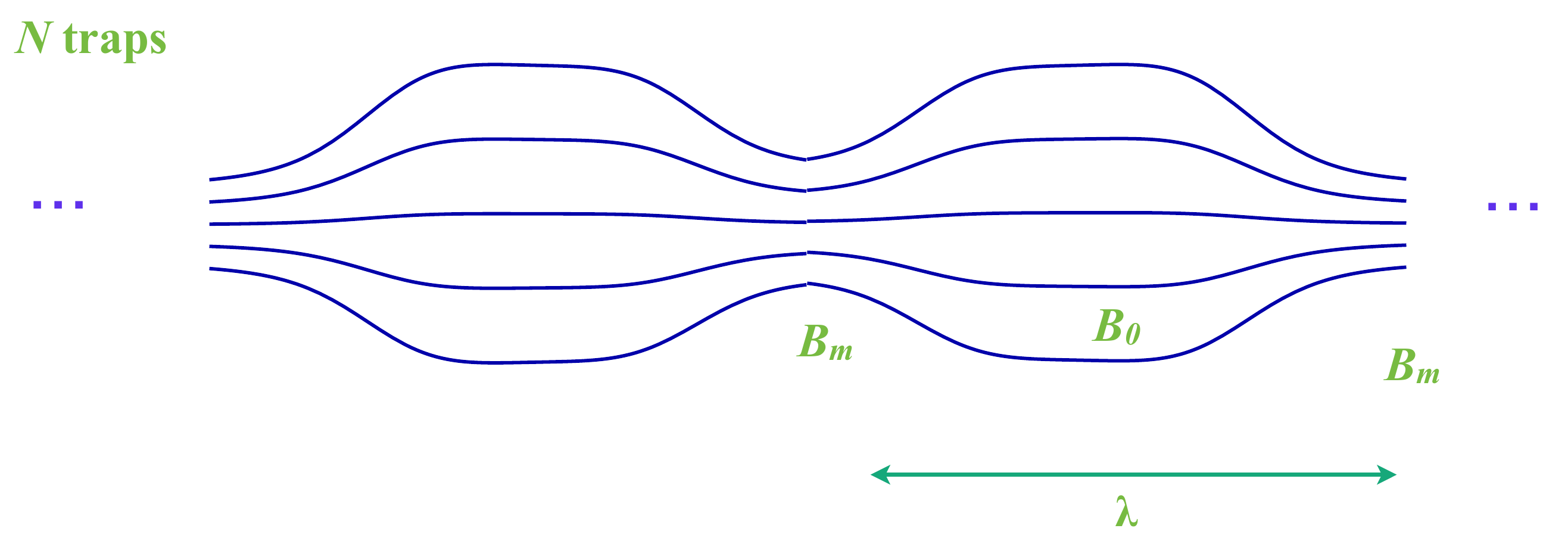}
\caption{A set of $N$ equivalent symmetric magnetic traps or ``magnetic bottles''. The minimum of the magentic field, $B_0$, is in the center, the maximum field, $B_m$, is at the magnetic mirrors on both ends of the bottle. }
\label{ntraps} 
\end{figure}

Here we first discuss the effect of magnetic trapping on the transport of particles using the example of a ``multiple-mirror machine'' illustrated in Fig. \ref{ntraps}. This model serves as a good toy model of statistically isotropic high-amplitude MHD turbulence in the ISM. Let us assume that the machine consists of a chain of $N\gg1$ identical magnetic traps, each of length $\lambda$, the low-field strength is $B_0$ and the mirror field is $B_m >B_0$; hence the mirror ratio is $R=B_m/B_0$. Efficient confinement requires that $r_L\ll\lambda$, so we assume it is satisfied. The angular size of the loss cones, $\theta$, is $\sin^2\theta=1/R$. 

Next, we assume that there is pitch-angle diffusion of particles, so they can be trapped and de-trapped in the magnetic bottles while they propagate from one trap to another. The most efficient regime is, of course, when the mean free path, $l_{pa}$, due to the pitch angle diffusion (i.e., the length over which the pitch angle changes appreciably, by about $\pi/2$) is comparable to $\lambda$. If otherwise $l_{pa}\ll\lambda$, then the system is highly collisional, so the effect of trapping is suppressed. Alternatively, if $l_{pa}\gg\lambda$, the particles can move through several traps in one ballistic run, before they become trapped again. This effectively increases the trap size and diminished the effect of trapping too. For the sake of argument, we assume that the rate of pitch-angle diffusion is optimal: $l_{pa}\simeq\lambda$. For cosmic rays in the galaxy, the pitch-angle diffusion coefficient is energy dependent, so does the above relation. Here we also introduce the effective collision time-scale $\tau\sim l_{pa}/v_{th}$, where $v_{th}\sim c$ is the thermal speed being of the order of the speed of light for CR, of course. 

Transport in such multi-mirror machines has first been studied elsewhere \cite{multi-mirror}. Obviously, such transport is diffusive: upon de-trapping from $i$-th trap, a particle will randomly go toward $(i+1)$-th or $(i-1)$-th one with 50\% chance. Next, we note that transport is the flux of streaming particles, i.e., the particles in the loss cones. Their fraction is about $\sim1/R$, therefore the number of streaming particles is $n_{stream}\sim n_0/R$, where $n_0$ is the total particle number. The particle flux is $\sim n_{stream} v_{th}\sim n_0 v_{th}/R$, that is, the effective plasma drift speed through the traps is $v\sim v_{th}/R$.  Finally, the effective diffusion coefficient through the traps is suppressed by a factor of $1/R^2$, namely
\beq
D_{traps}\sim v^2\tau\sim \frac{v^2_{th}\tau}{R^2}\sim \frac{D}{R^2},
\eeq
where $D\sim v^2_{th}\tau$ is the collisional diffusion coefficient in the absence of magnetic traps. We again mention that here we considered the most optimal case of $l_{pa}\simeq\lambda$, which can be written as $\lambda\lesssim v_{th}\tau/R\ll L$, where $L$ is the size of the whole system.

Using this result for CR propagation, we conclude that high-amplitude MHD turbulence present in the ISM suppresses their diffusion coefficient by a factor of the order of 
\beq
R^2\sim \left(\frac{\langle B \rangle +\tilde B}{\langle B \rangle } \right)^2, 
\eeq
that is by about an order of magnitude if $\tilde B/\langle B \rangle \sim2$ for example.

\section{Asymmetric diffusion}

\begin{figure}
\includegraphics[scale=0.3]{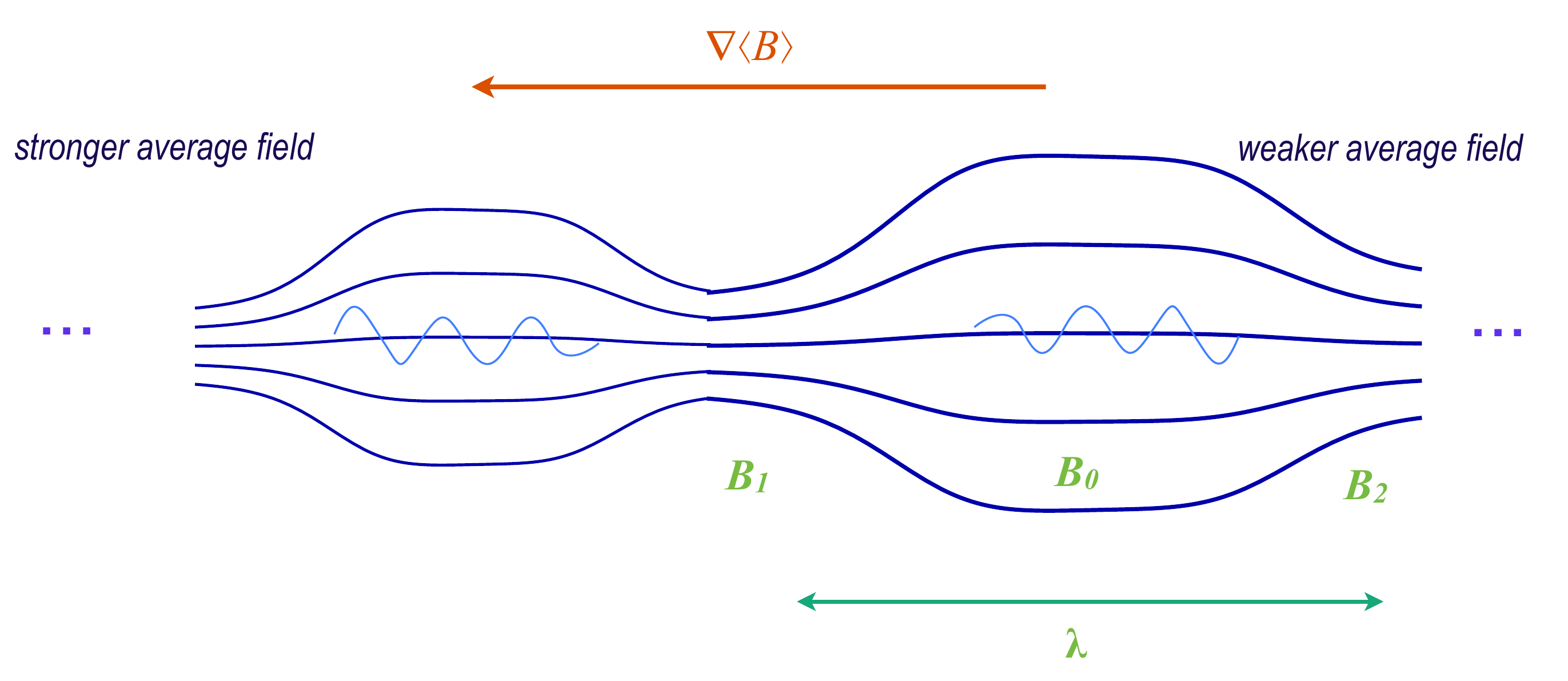}
\caption{The cartoon representing magnetic bottles in the presence of the mean field gradient. This is also a simplified one-dimensional model of the Galaxy. Here, the small waves in the trap centers stand for the low-amplitude Alfenic turbulence responsible for pitch-angle scattering of cosmic rays. The trap size, $\lambda$, is the effective correlation length on the outer scale of Alfvenic turbulence in the Galaxy. }
\label{bottles} 
\end{figure}

Now, we turn to the case with non-vanishing mean field gradient. For this we modify the previous multi-mirror machine toy model by imposing a long-scale variation of the global field. The magnetic traps become asymmetric, as shown in Fig. \ref{bottles}. This is the simplest, yet much more realistic one-dimensional model of the ISM, in which the global field strength has a gradient toward the galactic plane (i.e., to the left, in Fig. \ref{bottles}). The characteristic gradient scale is $H\sim1.5$~kpc, which is the observed galactic magnetic scale-height. The trap sizes are of the order of the outer scale of MHD turbulence. They vary across the Galaxy but typically of the order of $\lambda\sim1-10$~pc \cite{32}. 

\begin{figure}
\includegraphics[scale=0.3]{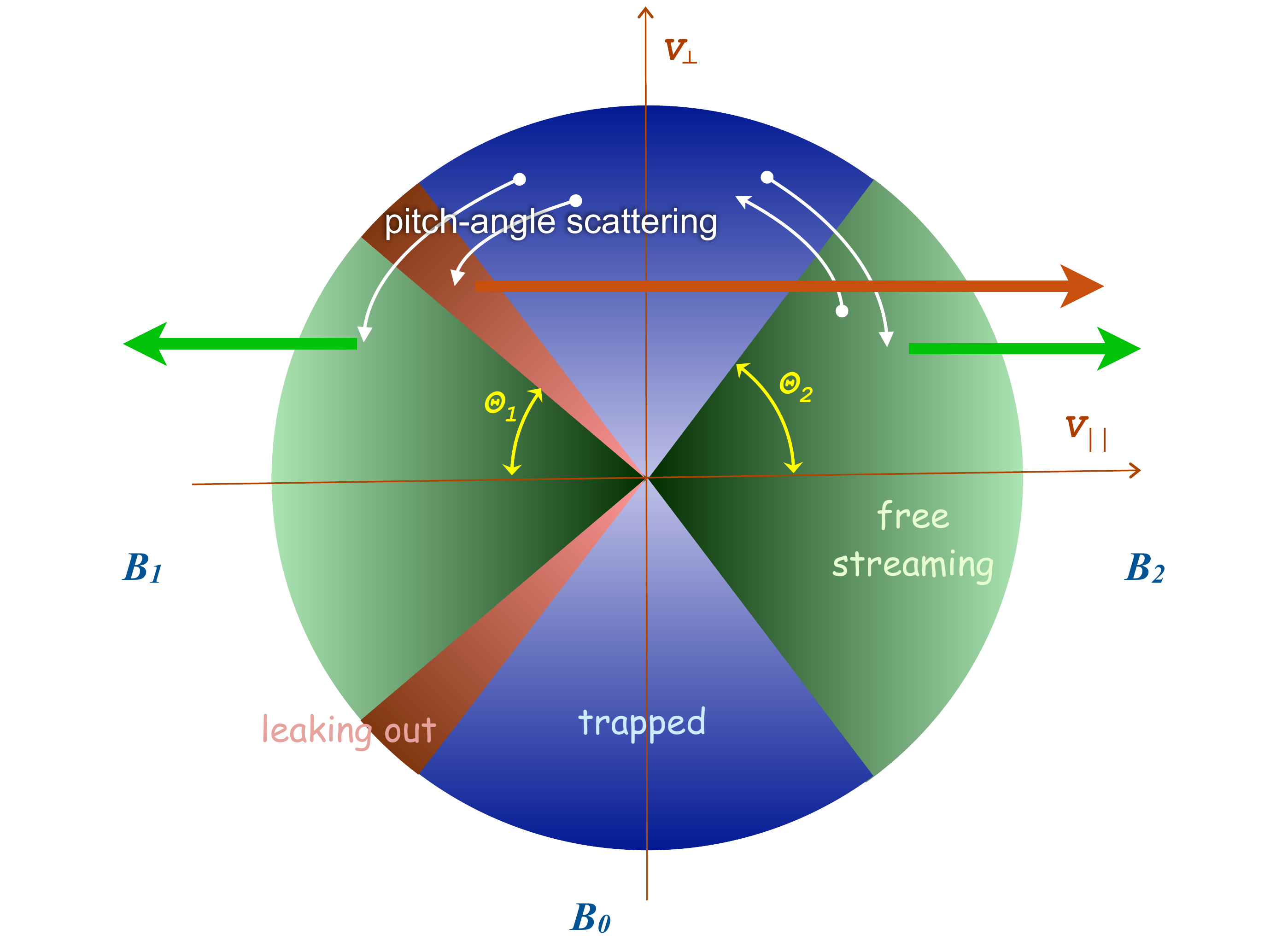}
\caption{The particle distribution function in a center of an asymmetric magnetic trap with $B_0\ll B_2<B_1$, where $B_0$ is the field at its center. Green wedges along $v_\|$ are the loss-cones with the corresponding opening angles $\Theta_1<\Theta_2$, the blue wedges are redions of trapperd particles, red wedges at $v_\|<0$ are the ``semi-trapped' particles representing the ``leaking out'' population-- they are reflected from the stronger field region on the left and then leak our from the trap through the right end. White arros represent the effect of the pitch angle diffusion which traps and de-traps particles inside the trap. }
\label{losscones} 
\end{figure}

The most important difference with the previous model is the slight asymmetry of the mirrors of individual traps, which leads to the substantial difference in the particle distribution function, see Fig. \ref{losscones}. An asymmetric magnetic trap has mirror fields $B_1>B_2\gg B_0$, where $B_0$ is the field at its center. Correspondingly, the loss cones differ too, $\theta_2>\theta_1$. The loss cones, shown in green, correspond to the streaming population of particles. The trapped population (shown in blue) is, in contrast, symmetric. Finally, the red population are the particles moving to the left. They will be reflected from the left mirror and, in a half of the bounce time, will leak out from the trap through the weaker right mirror. White arrows illustrate pitch angle diffusion. 

Obviously, the particle populations escaping through both ends are not equal and proportional to the volumes of the spherical cones of the with the opening angles, $\theta_{1,2}$, namely

\bea
\Omega_1&=&\left(\frac{2\pi}{3}\right)\left(1-\cos\theta_1\right)
\nonumber\\
&=&1-\left(\frac{\Delta B_1}{B_1}\right)^{1/2},
\\
\Omega_2&=&\left(\frac{2\pi}{3}\right)\left[(1-\cos\theta_2)+(\cos\theta_1-\cos\theta_2)\right]
\nonumber\\
&=&1+\left(\frac{\Delta B_1}{B_1}\right)^{1/2}-2\left(\frac{\Delta B_2}{B_2}\right)^{1/2},
\eea
where $\Delta B_1=B_1-B_0$ and $\Delta B_2=B_2-B_0$ and we used that $\cos\theta=(1-\sin^2\theta)^{1/2}=(1-B_0/B)^{1/2}=\left(\Delta B/B\right)^{1/2}$.

\begin{figure}
\includegraphics[scale=0.3]{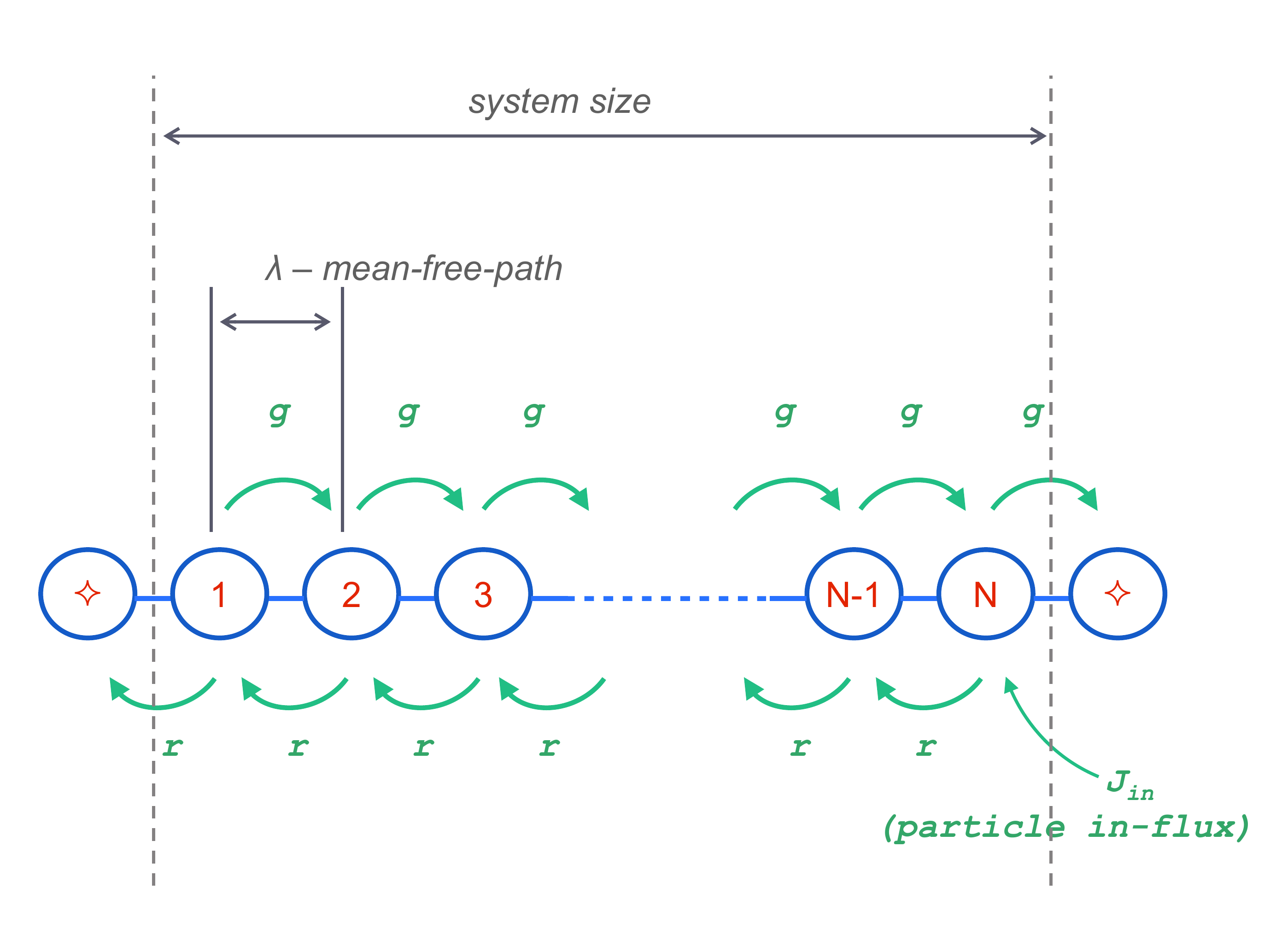}
\caption{The Markov chain representing the toy model of one-dimensional asymmetric diffusion. The transition probabilities to the left, $r$, and to the right, $g$, are constants but generally $r\not=g$. The in-flux of particles, $J_{in}$, into the system occurs through the right boundary. Each cell represents a magnetic trap; there separation is of the order of the cell size, $\lambda$, which plays a role of the effective mean free path. }
\label{chain} 
\end{figure}

To proceed further we introduce a mathematical model of a one-dimensional Markov chain \cite{soc96,soc98}. It is shown in Fig. \ref{chain}. Here, every magnetic trap is represented by a site, where a particle stays for some collision time, $\tau$, before it de-traps and rapidly jumps left or right into a neighboring site. As for any diffusive process on a Markov chain, we assume for simplicity that such de-trapping/re-trapping events occur at regular intervals, $\tau$. The number of sites is $N=L/\lambda$ with $L$ being the system size. In order to account for the leakage from the system, we have to add two ``absorbers'' at both ends of the chain --- the $*$-states, often referred to as the ``limbo''-states. Finally, we have to have a source of particles. For example and without loss of generality, we  allow for the in-flux of particles, $J_{in}$, through the right boundary.

The transition probabilities at each time step are $r$ and $g$, respectively. Standard diffusion is known to have equal probabilities to go left or right --- this is the classical mathematical problem, the ``drunken sailor problem''. In asymmetric diffusion, in contrast, the probabilities of the forward and backward
transitions are not equal. In general, these probabilities may be different for different sites. Here we assume that they are constant throughout the chain. The values of $r$ and $g$ are the fractions of the left- and right-streaming particles, respectively, that is they are $r=\Omega_1/(\Omega_1+\Omega_2)$ and $g=\Omega_2/(\Omega_1+\Omega_2)$. It is convenient to cast them into the form:
\beq
r=1/2-\epsilon, \quad g=1/2+\epsilon,
\eeq
where
\beq
\epsilon=\frac{\delta B}{4B_2}\frac{B_0}{\left(B_2\,\Delta B_2\right)^{1/2}}
\left[1-\left(\frac{\Delta B_2}{B_2}\right)^{1/2}\right]
\eeq
and $\delta B=B_1-B_2$. Since the mean field varies only slightly on the scale of a single trap, we can estimate that
\beq
\delta B\sim\lambda|\nabla B|\sim (\lambda/L) B_0.
\eeq
The factor $\lambda/L$ can be interpreted as the ratio of the turbulence outer scale to the characteristic mean field gradient scale. We can also estimate that $\Delta B_2=(B_2-B_0)\sim \tilde B$ and $B_1/B_0\simeq B_2/B_0\sim R$, so that 
\beq
\epsilon\sim\frac{1}{8R^2}\,\frac{\lambda}{L}.
\eeq
Note the appearance of the $1/R^2$ suppression factor akin the one obtained in the previous section. 

The dynamics of a Markov chain is determined by the ``continuity equations'' written for each site. They are 
\bea
d_t n_1 &=& - (r_1 +g_1 )n_1 + r_2 n_2, \label{left} \\
d_t n_2 &=& - (r_2 +g_2 )n_2 + g_1 n_1 + r_3 n_3, \label{second}\\
& & \dots \nonumber \\
d_t n_j &=& - (r_j +g_j )n_j + g_{j-1} n_{j-1} + r_{j+1} n_{j+1}, \label{inner}\\
& & \dots \nonumber \\
d_t n_N &=& - (r_N +g_N )n_N + g_{N-1} n_{N-1} +J_{in}, \label{right}
\eea
where $d_t$ stands for the time derivative and $n_j$ is the probability of a particle to be at the $j$-th site. Note that the first and last equations describe the left and right boundaries of our system. For an ensemble of CR particles, $n_j$ is also the relative number density of particles in the $j$-th trap. Hereafter we assume the steady state, for simplicity, so that all time derivatives vanish, $d_t n_j =0$. 

\begin{figure}
\includegraphics[scale=0.5]{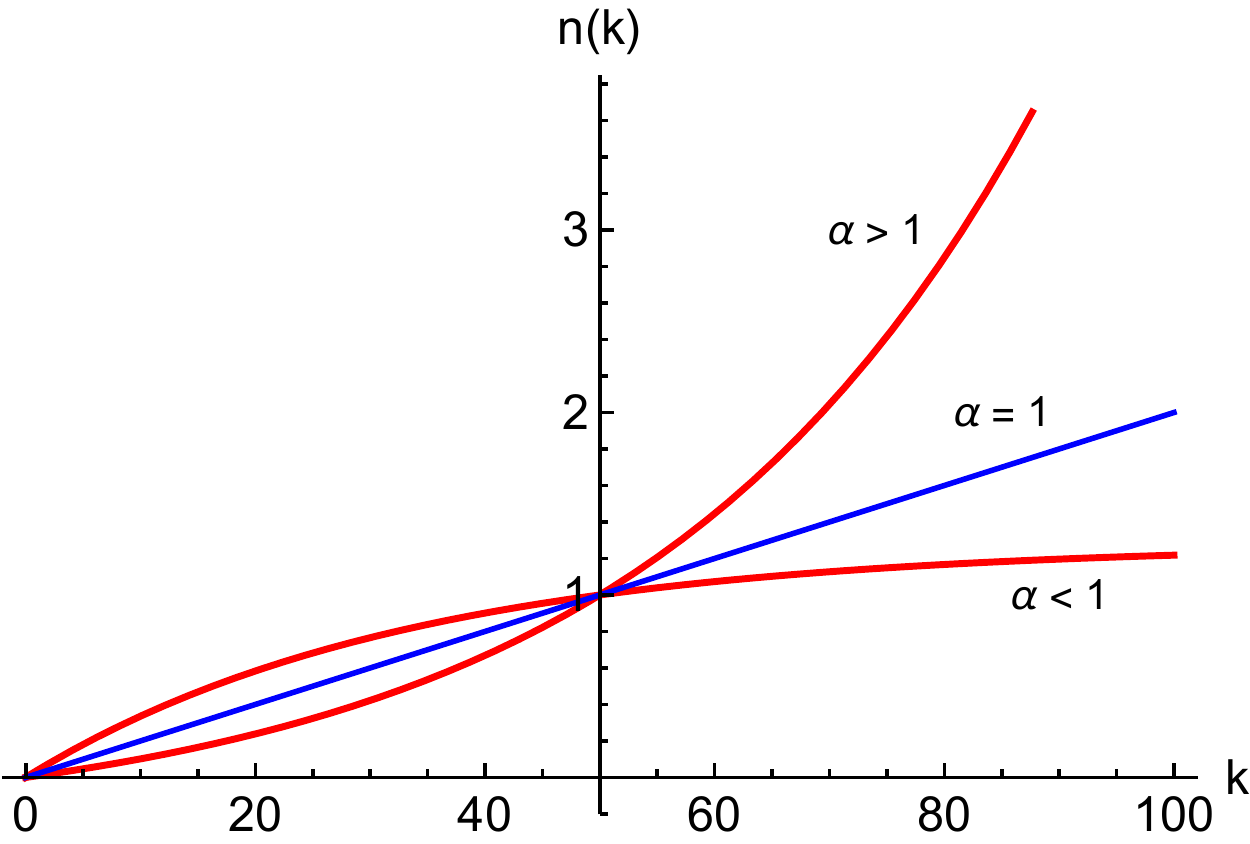}
\caption{The solution of the one-dimensional Markov chain model with constant coefficients $g$ and $r$. Asymmetric diffusion corresponding to the asymmetry parameter $\alpha\equiv g/r\not=1$ is shown by red curves. For comparison, the linear density gradiant predicted by the standard diffusion, $\alpha=1$, is shown with the blue curve. Here we took $N=100$ and normalized al the curves to the same value at $k=50$ to ease comparison.}
\label{dens} 
\end{figure}

We solve the system of $N$ coupled equations (\ref{left})--(\ref{right}) by calculating the {\em generating function} \cite{soc96,soc98} defined as follows
\beq
F(\xi)\equiv\sum_{j=0}^N \xi^jn_j,
\eeq
where $\xi$ is a dummy variable having no physical meaning whatsoever. Then, $n_k$ at some $k$-th site is simply the $k$-th derivative of $F(\xi)$ evaluated at $\xi=0$, namely
\beq
n_k=\frac{1}{k!}\left.\frac{d^k F(\xi)}{d\xi^k}\right|_{\xi=0}.
\label{deriv}
\eeq

The generating function is obtained, in turn, from the Markov chain equations as follows. One multiplies, the first equation by the first power of $\xi$, the second equation by $\xi^2$ and so on until the last equation being multiplied by $\xi^N$. Then we sum up all the resulting equations. Upon straightforward manipulations to isolate the combinations $\sum_{j=0}^N \xi^jn_j$, we obtain
\beq
F(\xi)=\frac{\xi(\xi^{N}J_{in}/r-\xi^{N+1}\alpha n_N-n_1)}{\left[(1+\alpha)\xi-1-\alpha \xi^2\right]},
\eeq
where we introduced the asymmetry parameter $\alpha=g/r$.

Using Eq. (\ref{deriv}), we obtain
\beq
n_k=(1+\alpha+\alpha^2+\dots+\alpha^{k-1})\,n_1=\frac{\alpha^k-1}{\alpha-1}\,n_1
\eeq
valid for $1\le k\le N$. Next, we apply the right boundary condition, namely that the $(N+1)$-th site is a limbo state, so $n_{N+1}$ is identically zero. Thus
\beq
n_{N+1}=-J_{in}+\frac{\alpha^{N+1}-1}{\alpha-1}\,n_1\equiv 0
\eeq
defines the yet unknown $n_1$. 

The final result --- the analytical solution of the set of Markov chain equations  (\ref{left})--(\ref{right}) --- is
\beq
n_k=J_{in}\left(\frac{\alpha^k-1}{\alpha^{N+1}-1}\right)\,,\quad1\le k\le N.
\label{soln}
\eeq
This result is presented in Fig. \ref{dens}, which shows the particle densities as a function of coordinate (to be precise, of the trap number, $k$). The salient feature of asymmetric diffusion is the non-linear particle density gradient. Depending on the value of $\alpha$, one sees exponential suppression/enhancement of the density. The linear gradient is recovered in the case of symmetric, i.e., standard, diffusion for which $\alpha=1$.

\section{Galactic example}

\begin{figure}[b]
\includegraphics[scale=0.3]{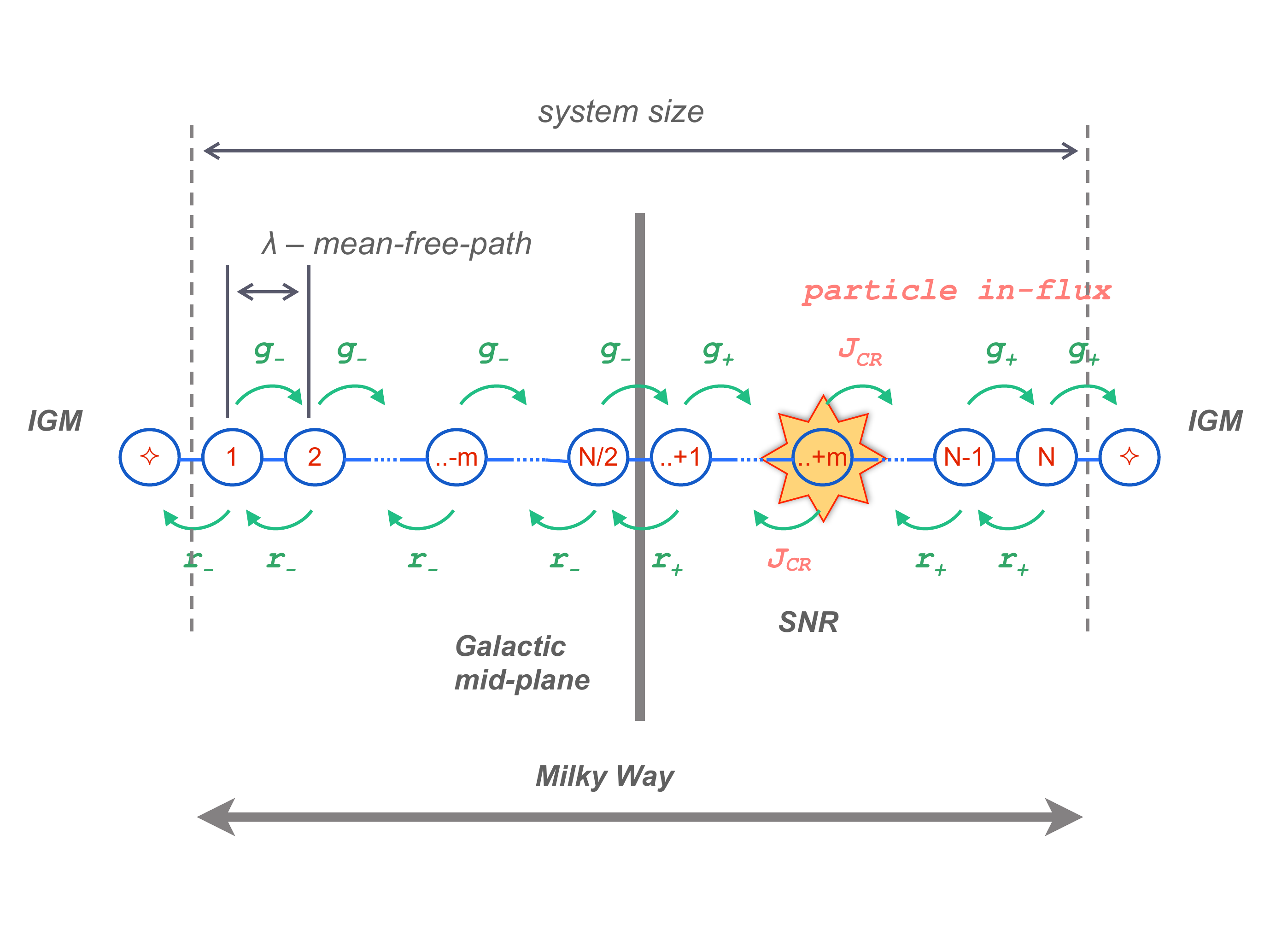}
\caption{The Markov chain representing CR diffusion in the Galaxy. The Galaxy is seen edge-on with the galactic plane being vertical. The galactic north is to the right. The system size is set by the galactic magnetic field scale-height, so the strongest field is at the mid-plane. Thus, the $g$ and $r$ probabilities are unequal on either side and, by symmetry, $g_+=r_-$ and $g_-=r_+$. The cosmic rays are injected by a supernova (SN) located at some $m$'th site to the right, i.e., at $k_{SN}=N/2+m$ site.}
\label{galaxy} 
\end{figure}
\begin{figure}
\includegraphics[scale=0.5]{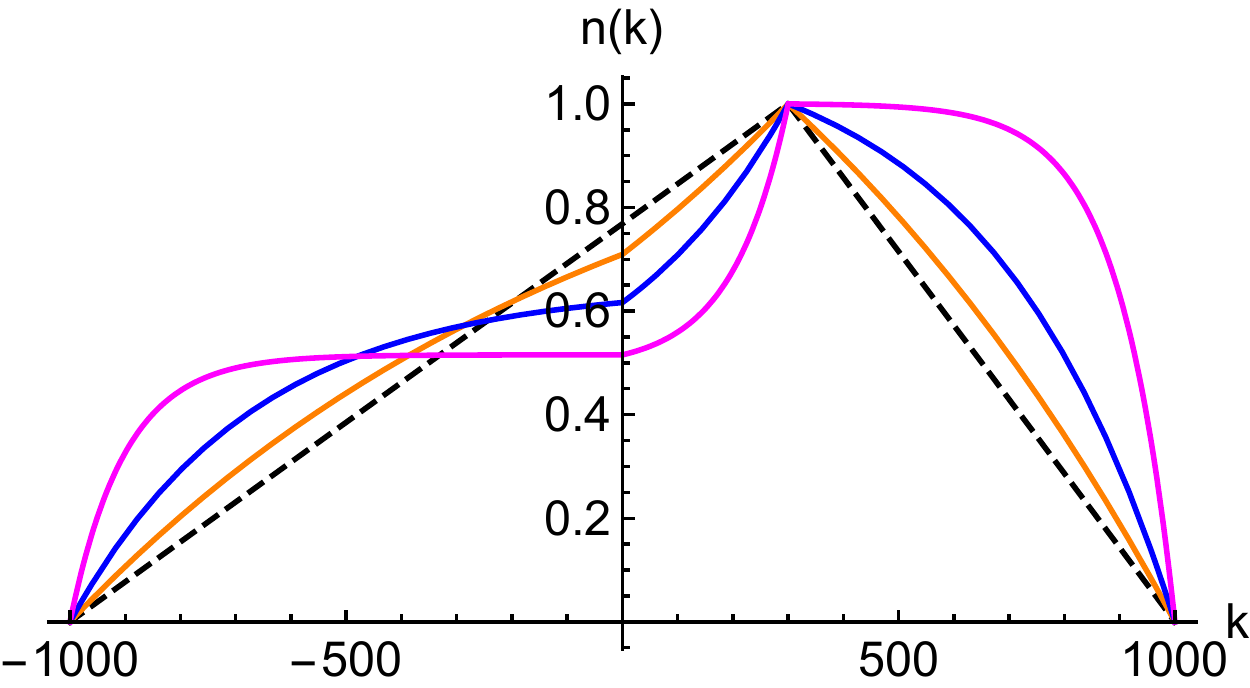}
\caption{The solution of the one-dimensional Markov chain model of our Galaxy with $N=2000$, and the cosmic ray source located 300 cells away from the Galactic mid-plane toward the galactic north (i.e., to the right), that is at $k_{SN}=1300$. The standard diffusion prediction ($\alpha=1$) is shown with the dashed line. The orange, blue and magenta curves represent the cosmic ray densities for asymmetric diffusion with $\alpha=1\pm\epsilon$ with $\epsilon=10^{-5},\ 10^{-3},\ 10^{-2}$ respectively.}
\label{crdens} 
\end{figure}

We now give an illustrative example pertinent to cosmic rays: the asymmetric CR diffusion in the direction orthogonal to the galactic plane. Fig. \ref{galaxy} illustrates the Markov chain for this system. The Galaxy is seen edge-on and the galactic plane is vertical and located between sites $N/2$ and $N/2+1$. The source of CR is a supernova located inside the galaxy at some site with the number $k_{NS}=N/2+m$, that is $m$-th site to the right from the plane. The CR particle flux, $J_{CR}$, is ejected in both directions. As we've mentioned earlier, the size of the system is set by the exponential scale height, $L=2H$ with $H\sim1.5$~kpc and the size of magnetic traps being comparable to the outer scale of ISM turbulence, $\lambda\sim1-10$~pc. Thus, we estimate the chain length to be $N\sim L/\lambda\sim1000$. 

The magnetic field is inhomogeneous: it decreases away from the galactic plane toward both ends. Therefore $r$ and $g$ probabilities differ on the left ($-$) and on the right ($+$) from the plane. By symmetry, $g_+=r_-$ and $g_-=r_+$.

This Markov chain can be solved analytically in the same way outlined in the previous section. Alternatively, one can use the generic solution, Eq. (\ref{soln}), to piece-wise construct the global solution by matching the solutions with $\alpha<1$ and $\alpha>1$ at the mid-plane and the SN location. The resulting equation is cumbersome but not instructive, so we just plot the solution in Fig. \ref{crdens}. This solution has $N=1000$,\ $k_{SN}=300$ and several $\alpha=1\pm\epsilon$ with $\epsilon=0,\ 10^{-5},\ 10^{-3},\ 10^{-2}$. Which one does correspond to our Galaxy? Assuming that the asymmetric diffusion-induced anisotropy should not exceed the observed CR anisotropy, which is about 0.1\%, we estimate $\epsilon$ to be $\epsilon\lesssim10^{-3}$. This corresponds to the blue curve, which shows about 50\% difference compared to the standard diffusion picture.

\section{Comment on a continuum limit}

It is instructive to determine how a finite-step-size diffusion on a Markov chain can be extrapolated to a continuum limit. To do this, we consider the generic Markov chain equation  
\beq
\frac{d n_j}{dt}=n_{j-1}g_{j-1}+n_{j+1}r_{j+1}-n_j(g_j+r_j)
\label{nj}
\eeq
and treat $j$ as a continuous variable --- a dimensionless distance, $x$, measured in the units of a diffusion mean-free-path, that is $\lambda$. Time is also dimensionless here because it is measured in the units of the collisional time-step, $\tau$. As long as $j\gg1$, one can Taylor expand in small $\Delta j=1\ll j$ and substitute $j$ with $x$, for convenience. Upon doing so, the transition term from the $(j-1)$-th site to the $j$-th results in the following substitution
\beq
n_{j-1}g_{j-1}
\to n(x)g(x)-\frac{d(ng)}{dx}+\frac{1}{2!}\frac{d^2(ng)}{dx^2}-\frac{1}{3!}\frac{d^3(ng)}{dx^3}+\dots
\eeq
Similarly, we treat the $(j+1)$-th site. 

Eq. (\ref{nj}) readily becomes, assuming $r$ and $g$ being constants,
\beq
\frac{\partial n}{\partial t}=(r-g)\frac{\partial n}{\partial x}+\frac{(r+g)}{2}\frac{\partial^2 n}{\partial x^2}+\frac{(r-g)}{3!}\frac{\partial^3 n}{\partial x^3}+\dots\ .
\eeq
The first two terms here are the constituents of the famous convection-diffusion equation
\beq
\frac{\partial n}{\partial t}=v\frac{\partial n}{\partial x}+D\frac{\partial^2 n}{\partial x^2} .
\eeq
Thus, the convection-diffusion equation is the expansion of the Markov chain equation up to the second order. Hence, it holds only when the higher order derivative terms are small enough. The drift (flow) velocity and the diffusion coefficient are identified as follows
\beq
v=r-g, \quad D=(r+g)/2.
\eeq
We also note that the density profile, Eq. (\ref{soln}), becomes, in the continuum limit and large $x$:
\beq
n(x)\propto e^{x\ln\alpha},
\eeq
if $\alpha>1$, that is exponential, and 
\beq
n(x)\propto 1-e^{x\ln\alpha},
\eeq
if $\alpha<1$.

\section{Conclusions}

We demonstrated that asymmetric diffusion of particles can occur in high-amplitude MHD turbulence with non-zero mean magnetic field gradient. The predicted density profiles are markedly different from the standard diffusion paradigm. We argued that accounting for asymmetric diffusion of CR in our galaxy can change their predicted density distribution by a factor of two. We also indicated that the familiar convection-diffusion equation of particle transport is just the first two terms of expansion of the more generic Markov chain equations. 

We stress that the asymmetric diffusion discussed here should not be mixed with anisotropic diffusion. The former deals with different probabilities of a particle to jump left or right, whereas the standard (even anisotropic) diffusion has those probabilities equal, though they may be position- and orientation-depend. 

Second important point to stress is that asymmetric diffusion results in a non-zero particle flux through the system, even though the plasma and the magnetic fields are completely static. This is in contrast with a conventional lore that the global (``bulk'') motion of CR in the galaxy may only be associated with global motions of the ISM, i.e., winds. This may particularly be important for the interpretation of the particle transport to/through the recently observed ``Fermi bubbles''.

\acknowledgements
This work has been supported by NSF grant AST-0708213 and DOE grant  DE-FG02-07ER54940.

\end{document}